# Data Therapist: Eliciting Domain Knowledge from Subject Matter Experts Using Large Language Models

SUNGBOK SHIN, Aarhus University, Denmark

HYEON JEON, Seoul National University, South Korea

SANGHYUN HONG, Oregon State University, USA

NIKLAS ELMQVIST, Aarhus University, Denmark

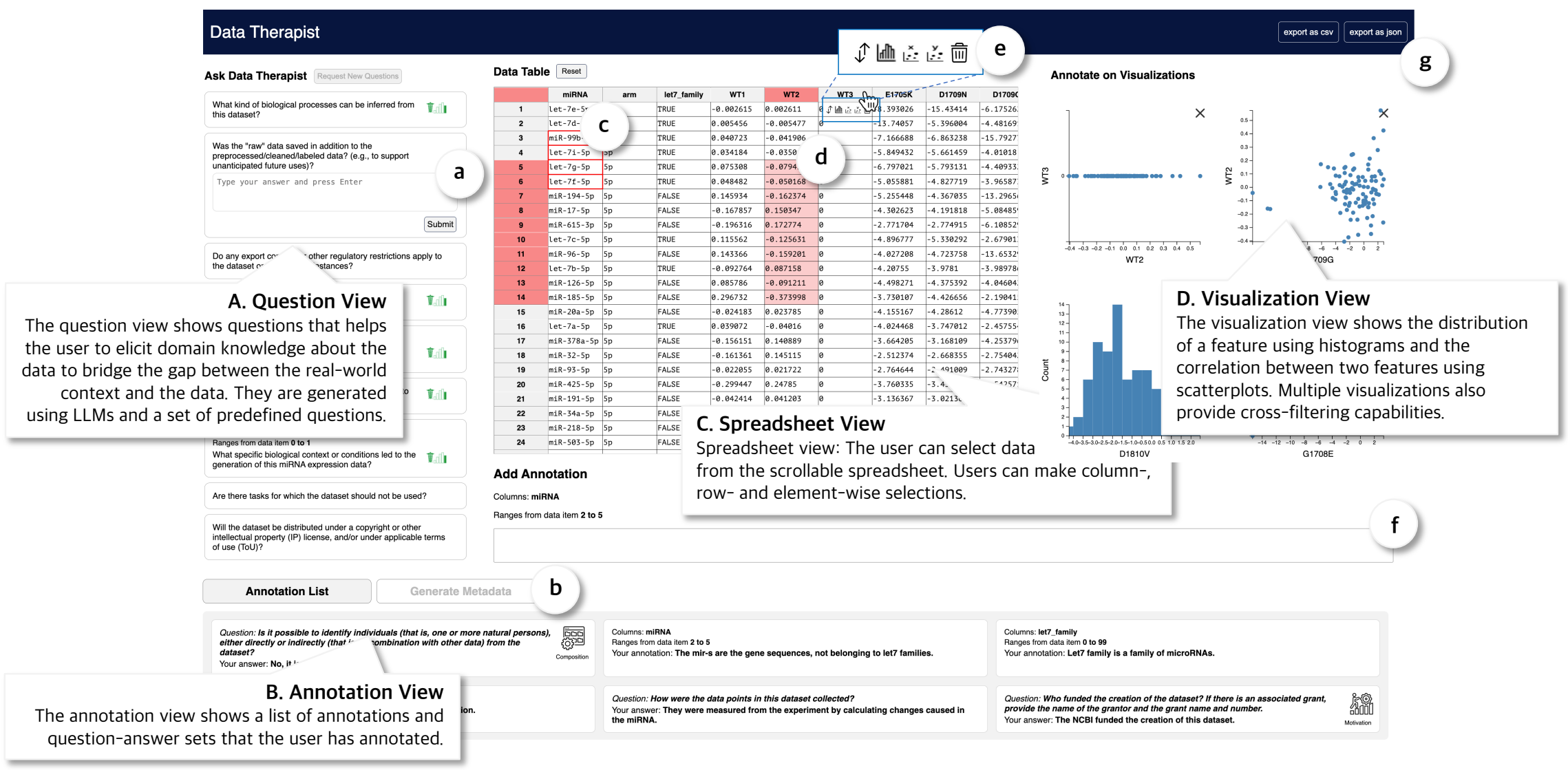

Fig. 1. Data Therapist User Interface is designed to elicit domain knowledge from experts about a domain-specific dataset by asking questions and capturing answers as annotations. These annotations bridge the gap between the real-world context and the data. Users (domain experts) can add annotations in three ways: (A) by answering questions in Question View, (B) by reviewing and editing annotations in Annotation View, and (C–D) by providing data-specific annotations through Spreadsheet and Visualization Views.

Effective data visualization requires not only technical proficiency but also a deep understanding of the domain-specific context in which data exists. This context often includes tacit knowledge about data provenance, quality, and intended use, which is rarely explicit in the dataset itself. Motivated by growing demands to surface tacit knowledge, we present the Data Therapist, a web-based system that helps domain experts externalize such implicit knowledge through a mixed-initiative process combining iterative Q&A Authors' Contact Information: Sungbok Shin, Aarhus University, Aarhus, Denmark, sbshin90@umd.edu; Hyeon Jeon, Seoul National University, Seoul, South Korea, hj@hcil.snu.ac.kr; Sanghyun Hong, Oregon State University, Corvallis, USA, sungbok.shin@inria.fr; Niklas Elmqvist, Aarhus University, Aarhus, Denmark, elm@cs.au.dk.

with interactive annotation. Powered by a large language model, the system automatically analyzes user-supplied datasets, prompts users with targeted questions, and supports annotation at varying levels of granularity. The resulting structured knowledge base can inform both human and automated visualization design. A qualitative study with expert pairs from Accounting, Political Science, and Computer Security revealed recurring patterns in how expert reason about their data and highlighted opportunities for AI support to enhance visualization design. All supplemental materials are available at https://osf.io/e4fxu.

## 1 Introduction

*"Tell me about your data."* This question, or some variant thereof, is one that most data visualization experts will reach for whenever they start a new project in a new domain. As visualization researchers, our business is data first and foremost, but the data we visualize is invariably someone else's. And so, as part of this exploratory process to gather necessary information about a new domain—a process that has variously been referred to as design study methodology [39], user-centered design [13, 53], or participatory design [33] depending on your background and approach to visualization—we tend to find ourselves engaging with the owners of this data in interviews that are not unlike what a psychotherapist will engage in with a patient. Regardless of domain, the questions we ask will be more or less similar: the source of the data, how often it is updated, its quality, the stakeholders responsible for it, and so on. Nevertheless, understanding the data model of a domain, as well as its metadata and tacit knowledge, is a vital activity in designing a visualization that addresses user needs.

In this paper, we present the Data Therapist, a web-based system designed to elicit domain knowledge from subject matter experts. Specifically, we focus on capturing the deeper tacit knowledge that experts hold about their data—knowledge not addressed by existing approaches that enhance transparency and documentations, such as datasheets [12]. Inspired by mixed-initiative interaction [15, 18], Data Therapist combines iterative question-answering with user annotation. By leveraging large language models (LLMs), the system automatically generates knowledge-eliciting questions that prompt experts to articulate explanations and context, which are then captured as annotations. This structured externalization of tacit knowledge supports more effective communication, particularly in scenarios where subject matter experts (our users) and visualization designers cannot collaborate directly.

Data Therapist automates this process using guided question-and-answer dialogue to elicit domain knowledge and a direct annotation interface to help experts' annotations. Starting with a dataset uploaded by the domain expert, the Data Therapist will analyze the data and ask a mix of questions based on the seven genres of metadata [12] as well as its growing body of knowledge about the data and the domain. Experts can choose between answering questions, annotating the overall dataset, or adding data-specific annotations using their domain knowledge. The Data Therapist system operates in the following manner (Figure 1): to begin with, it gathers expert-driven explanations about the dataset of interest in the form of annotations. Based on the annotations, the LLM generates follow-up questions that deepen understanding of the topic, whether by clarifying underlying mechanisms or unpacking difficult terminology. The answers that the expert chose to answer from these questions are then added to the list of annotations. Data-specific annotations can either be added to columns, rows, or individual cells in an interactive spreadsheet, or to interactive visualizations that are automatically created as the user selects data. All answers are automatically added and organized into a structured knowledge base that can be exported at the end of an elicitation session and shared with a visualization designer—or even used as the basis for an automated visualization recommendation system.

We have validated the Data Therapist through a web-based implementation that uses a server-side Large Language Model (GPT-4 in our case) to formulate questions, structure elicited data, and provide contextual memory for guiding further inquiries into the problem domain. Furthermore, we use the Data Therapist in a qualitative user study involving four sets of two domain experts drawn from three distinct problem domains: computer security, political science, and accounting. We evaluate the capability of LLMs by comparing this with that of a human annotator, and understand what DT is capable of, as well as to explore the challenges. Our two-phase evaluation protocol begins by asking the first domain expert in each pair to use Data Therapist for a 45-minute session to elicit domain knowledge. Simultaneously, the second expert performs the same task with the help of a human annotator, instead of Data Therapist. In the second phase, the two experts share their results and assess the quality of the annotations. We also analyze their usage patterns to understand how experts engage with the tool.

Our findings show that experts produce more annotations when using Data Therapist than when working with a human interviewer. On average, experts also rated Data Therapist as 5 percentage points less difficult to use than a human annotator. Qualitative analysis further reveals a challenge in juxtaposition with human annotator: that while Data Therapist-driven annotations are effective in clarifying dataset mechanisms, they are less successful at anticipating the details that laypeople might struggle to understand.

## 2 Background and Motivation

Our research intersects four areas of prior work: domain knowledge elicitation, metadata practices, LLM-driven knowledge management, and visualization design support. We review each in turn, identifying knowledge gaps and highlighting how our contributions address them.

### 2.1 (Domain) Knowledge Elicitation in Visualization Design

Effective visualization design requires both technical visualization expertise and deep domain knowledge. While several theoretical models describe how data transforms into visualizations [21, 34, 39, 45], these models often assume ideal knowledge transfer between domain experts and visualization designers.

In practice, professional visualization design typically involves a division of roles: domain experts provide data and define goals, while visualization designers create the visual representations [49]. This division exists because domain experts generally lack design expertise, time, and often motivation to create effective visualizations themselves [14, 46]. *However, it introduces communication gaps that can impede the design process*. Bigelow et al. [2] observed that "the designers' inferences about data behavior were often inaccurate" and "their inferences about the data structure are also separated from the actual data structure." Similarly, Walny et al. [46] noted that "for designers, anticipating all of the implications of scale and interaction for a given dataset remains challenging." Such findings highlight a fundamental problem: visualization designers lack crucial contextual knowledge that domain experts hold about their data.

This knowledge gap encompasses data provenance, quality, implicit assumptions, and domain-specific interpretations. Lin et al. [23] conceptualized this tacit knowledge as "data hunches"—insights about mismatches between reality and data representation. These hunches acknowledge that data is always an imperfect representation of reality. Capturing them at the data level, before visualization design begins, can significantly improve the quality of the final visualization [24].

Existing elicitation approaches, such as interviews [39, 53] and collaborative design sessions [33], are valuable but limited: they rely on unstructured conversations that may miss critical insights, and they often generate unstructured documentation that is difficult to systematically incorporate into visualization design decisions. This gap motivates the need for a *systematic* approach to elicit, structure, and apply tacit domain knowledge in visualization design.

*Our contribution.* Our work proposes a structured approach to eliciting and documenting tacit domain knowledge that addresses communication gaps between subject matter experts and visualization designers. Implemented in the Data Therapist system, our approach integrates LLM-driven questioning with interactive annotation to systematically capture “data hunches" and transform them into reusable knowledge. The structured externalization reduces the risk of misinterpretation and provides designers with contextual insights that improve the effectiveness of visualization.

### 2.2 Metadata and Documentation for Datasets

As datasets grow in complexity and importance, documenting their context, provenance, and limitations has become increasingly vital. Gebru et al. [12] proposed the concept of “datasheets for datasets,” a framework inspired by standardized documentation in the electronics industry. Their approach organizes metadata into seven genres: motivation, composition, collection process, preprocessing, uses, distribution, and maintenance. The goal is to improve transparency and accountability in data-driven systems.

In visualization research, metadata has also been studied as a way to improve interpretation, understanding, and communication. Burns et al. [4, 5] showed that disclosing metadata can strongly influence how audiences perceive reliability, particularly in decision-making contexts. Others have proposed incorporating metadata directly into visualization systems: Srinivasan et al. [42] introduced “data facts”—concise, automatically generated statements about statistical features or contextual information. These data facts augment visualizations to facilitate interpretation and communication, addressing challenges that arise when viewers lack domain knowledge or analytical expertise. The visualization of metadata itself has also emerged as a research focus. Moreover, Makhija et al. [28] developed Loch Prospector, a system for visualizing metadata for data lakes. Their approach helps users navigate large open data repositories through metadata representations, demonstrating how metadata visualization can aid understanding.

Despite these advances, current approaches remain limited in three ways: (1) Most documentation practices are *manual*, making them difficult to scale; (2) Documentation frameworks capture only *explicit characteristics* of datasets, overlooking the tacit domain knowledge experts hold that strongly shapes interpretations; and (3) Metadata documentation and visualization design are typically treated as *separate process*, leading to visualization that lack critical contextual factors known to domain experts but not formally recorded in the dataset or its documentation.

*Our contribution.* We addresses these limitations with Data Therapist, an interactive, mixed-initiative system for eliciting and documenting dataset metadata. Building on the seven genres of metadata proposed by Gebru et al. [12], our approach captures comprehensive dataset context and the tacit knowledge of domain experts. By generating a structured knowledge base that integrates directly with visualization design, Data Therapist bridges the gap between manual documentation practices and design workflows. LLM-guided questioning and annotation further reduces effort required for through documentation while enabling experts to contribute their insights effectively.

### 2.3 (Large) Language Models as Knowledge Bases

A language model is a statistical model trained on a large corpus of text data to support *next token prediction*. A token is typically a word or subword, depending on how the model is designed. Given a sequence of tokens, the model computes the probability distribution over possible next tokens and chooses the most likely one.

Recent work has demonstrated that *scale* is the key factor in enabling them to accomplish complex tasks effectively [3, 11, 48]. These "large" language models (LLMs) typically contain billions of parameters [1, 3, 7, 35, 44], but the scale also makes training computationally intensive and resource-demanding. As a result, training is almost exclusively done by large-budget organizations such as OpenAI, Google and Meta, with the resulting *pre-trained* models offered either as online services (e.g., OpenAI's ChatGPT or Google's Gemini) or as open-source releases (e.g., Meta's LLaMA models).

Because LLMs are trained on massive corpora of curated text with large quantities of *dialogue and chat data*, they are not only proficient at diverse tasks but also naturally suited for interactive settings. This training regime equips them to engage in multi-turn conversations, generate questions, interpret answers, and adaptively follow-up—capabilities that make them particularly well-matched for eliciting domain knowledge through dialogue with subject matter experts.

Prior work has even examined LLMs as knowledge bases. Heinzerling and Inui [16] analyzed their entity representations, storage capacity, and question-answering abilities, demonstrating that LLMs can indeed store and retrieve factual knowledge, albeit with limitations in consistency and reliability compared to traditional structured knowledge bases. These limitations often become most visible in interactive use, where LLMs may produce inconsistent or speculative responses when generating and answering questions. To improve factual accuracy and reduce such inconsistencies in our setting, we follow the standard practice in the community [31]—designing prompts that explicitly guide the reasoning process of LLMs. By combining their conversational strengths with prompt-based scaffolding, our system leverages LLMs as effective partners for eliciting tacit knowledge in a structured and reliable manner.

*Our contribution.* To the best of our knowledge, this is the first work to leverage LLMs as active partners—rather than static repositories—in the knowledge elicitation process. Data Therapist uses an LLM to generate contextually relevant questions, interpret free-text responses, and organize elicited knowledge into a structured format aligned with established metadata frameworks. This structured dialogue enables domain experts to externalize tacit knowledge, addressing long-standing challenges in capturing and documenting it systematically. While not a substitute for visualization expertise, it bridges the gap between domain knowledge and visualization design by producing a reusable knowledge base that designers can directly leverage. By combining the conversational strengths of LLMs with structured annotation, our approach offers a more accessible and comprehensive alternative to manual documentation.

### 2.4 Augmenting the Visualization Design Process

The goal of automated visualization recommendation is to reduce the effort required to author effective visualizations by automatically suggesting designs that match data characteristics and user needs. Since Mackinlay's pioneering work [26, 27], this field has advanced significantly. Modern systems draw on diverse inputs, such as data properties [20, 50, 51], perceptual principles [50], expert feedback [25], and formalized design knowledge [32]. Recent AI-driven approaches [10, 19, 43, 55] extend this process by incorporating both data properties and usage context.

Beyond recommendation, design feedback mechanisms augment visualization authoring. Systems inspired by linting metaphor provide automated critiques of visualization code or output [6, 17, 30], while others leverage perceptual signals such as visual salience and color usage to inform design decisions [40, 41]. VisLab [9] introduces a crowdsourcing approach to gather diverse user feedback on visualization design.

Authoring tools likewise have focused on enhancing expressivity and flexibility. Systems such as Charagraph [29] and ChartSpark [52] support sophisticated customization and narrative intent, while Data Formulator [47] bridges natural language and visual representation by binding textual concepts to visualizations.

More recently, researchers have addressed issues of trust and transparency by incorporating data context into visualizations. Burns et al. [4, 5] show that metadata disclosure influence user trust and understanding, highlighting that effective visualization requires not just technical design knowledge but also a deeper understanding of data provenance and limitations.

Despite these advances, current recommendation and authoring tools—automated or otherwise—do not focus on *eliciting domain knowledge from users*, even through such knowledge is critical for the design process [39].

*Our contribution.* The Data Therapist addresses this gap by providing a structured framework for capturing and communicating domain experts' implicit knowledge. Unlike existing augmentation approaches that focus primarily on the technical aspects of visualization design, our system targets the upstream challenge of knowledge elicitation and documentation. This framework complements current recommendation and feedback systems by supplying the crucial contextual information needed to support more appropriate and effective design decisions.

## 3 Designing the Data Therapist

We introduce Data Therapist, a mixed-initiative [15, 18] annotation system designed to elicit metadata and tacit knowledge about datasets. To our knowledge, no prior work directly address this objective; existing approaches typically rely on predefined questions that enrich metadata only in limited ways. The Data Therapist advances beyond these baselines by (1) generating questions tailored to the dataset and its intended audience, (2) allowing flexible references to specific dataset elements, and (3) supporting in-situ analysis so that experts can contribute rich, contextual annotations Inspired by the role of a human therapist, Data Therapist engages domain experts in structured dialogue, combining directed questioning with freeform annotation. This dual approach enables experts to externalize both explicit and implicit knowledge, including insights that may seem obvious to them but are often invisible to non-experts. The following subsections detail the design considerations, conceptual model, and methods underlying Data Therapist.

### 3.1 Design Considerations

The Data Therapist's design is guided by several key design considerations (DCs) that shape its functionality and user experience:

- **DC1 Bridge reality and representation.** The system must elicit annotations that minimize the gap between reality and its data representation. Questions should prompt domain experts to explain context at a level that allows non-experts to understand the dataset's limitations, assumptions, and real-world significance.
- **DC2 Support multi-granular annotation.** Users should be able to flexibly select and annotate data at multiple levels of granularity—entire columns or rows, individual cells, or a combination—depending on where domain knowledge applies.
- **DC3 Enable analysis.** The system must incorporate analytical capabilities that allow experts to identify and explain patterns and anomalies in their data, facilitating the externalization of insights that might otherwise remain implicit.
- **DC4 Ensure quality.** The tool should provide mechanisms to verify annotation quality, identify contradictions between annotations, detect incomplete responses, and prompt for clarification or additional detail when necessary.

**DC5 Provide oversight.** Users need clear visibility into their annotation progress, including what aspects of the data have been addressed and what remains unexplained, enabling systematic and comprehensive knowledge capture.

**DC6 Facilitate comprehensive coverage.** The system should guide users toward complete domain knowledge capture by systematically addressing all relevant metadata categories. This includes proactively identifying knowledge gaps and steering the elicitation process toward unexplored aspects of the dataset.

### 3.2 Conceptual Model

The overarching goal of our work is to bridge the gap between data and real-world context [24] by eliciting knowledge from domain experts (DC1). Here the Data Therapist plays the role of a non-expert—it has access to the data but lacks background knowledge about its problem domain. To address the design considerations, the approach provides two complementary modes seamlessly combined in a mixed-initiative interaction [15, 18] model:

(1) A **guided question-and-answer dialogue** that probes for specific aspects of data context, and
(2) A **direct annotation interface** where experts can attach observations to the data at various levels of data granularity.

By interweaving these modes, the Data Therapist helps users progressively build a comprehensive knowledge base that captures both formal metadata and tacit domain insights. The purpose is not to shift reliance onto the system by having experts merely answer its questions, but rather to enhance the annotation process by asking questions that complement aspects the expert may not have initially considered.

### 3.3 Addressing the Design Considerations

The Data Therapist translates our design considerations into practical functionality through a structured mixed-initiative annotation workflow. When a domain expert begins with an unannotated dataset, the system activates complementary pathways for knowledge capture:

**For dataset-wide context (DC1, DC6):** Experts can add global annotations without selecting specific data points. These annotations typically cover provenance, collection methodology, and intended use cases as the foundational reality behind the datasets.

**For data-specific insights (DC2):** The system enables multi-granular selection, allowing experts to highlight individual cells, entire columns or rows, or arbitrary regions to annotate. This flexibility accommodates different types of domain knowledge that may apply at varying levels of specificity.

Throughout this process, the Data Therapist actively participates by:

- Generating targeted questions based on observed data patterns and metadata frameworks (addressing DC3, DC6);
- Cross-referencing new annotations against existing ones to flag potential inconsistencies (addressing DC4);
- Tracking annotation coverage across the dataset and metadata categories (addressing DC5); and
- Suggesting areas for further exploration based on detected knowledge gaps (addressing DC6).

This bidirectional interaction creates a knowledge elicitation loop where expert-initiated annotations inform system-generated questions, which in turn can prompt deeper expert insights. The resulting knowledge base captures both explicit metadata and tacit domain knowledge in a structured format that can inform visualization design decisions.

## 4 System Implementation

We describe how Data Therapist is implemented to realize the design decisions outlined in the previous section.

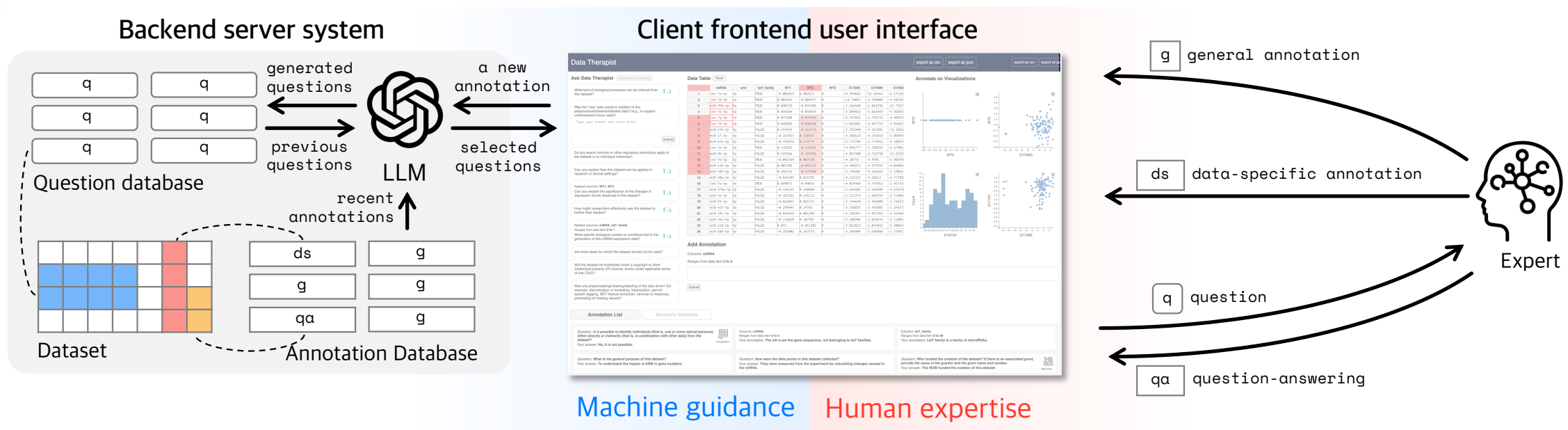

Fig. 2. **The Data Therapist overview and workflow.** The system consists of a user interface and a backend. Through the interface, domain experts provide both general and data-specific annotations, which are stored by the backend. The backend then generates additional questions based on newly added or existing annotations and presents them back to the user through the interface.

### 4.1 System Overview

The Data Therapist system adopts a server-client architecture to streamline data annotation and analysis. This design not only supports current functionality but also provides the flexibility to incorporate additional metadata types, richer annotations, and emerging LLMs, enabling the system to evolve and improve over time. An overview of the system is shown in Figure 2. The client provides the annotation and question-answering interface, while the server stores annotations and generates questions via an LLM. We present the user interface in § 4.2, followed by the three annotation options: general annotation (§ 4.3, data-specific annotation (§ 4.4), and question-answering (§ 4.5)). We then describe the question generation mechanism (§ 4.6; DC1, DC4) and the prompting methodology in § 4.7.

*Client frontend.* The client frontend interface provides functionality to users to elicit their domain knowledge about the dataset. First, it allows users to add annotations to the dataset and to help the process, it offers questions to guide towards enriching the annotation. It also offers multiple methods for selecting data instances within the dataset, such as spreadsheet-style views and feature-distribution visualizations (DC2, DC3) (see § 4.2). Users can directly make annotations. The user can also choose to make annotations by answering the questions that Data Therapist generates.

The system helps users track their progress by displaying a list of annotations, along with a thematic summary of the information they contain (DC5). Once a sufficient number of annotations have been generated, users can either export the full list of annotations or generate a dataset report based on the annotation-driven analysis.

Finally, to export the annotations (Figure 1g), the user can click the button 'export annotations' to receive a JSON file of all the annotations.

*Server backend.* The primary goal of the server is to manage user annotations and generate questions that elicit user responses about the dataset. Each time a user uploads an annotation, it is stored on the server. In addition, the server performs background operations to update the system—such as adding and prioritizing new questions, or generating summaries. These updates are processed on the server before being reflected in the client interface. We detail the prompting methodology in § 4.7. Finally, to support ongoing work, the server continuously saves the most recent version of the annotated dataset.

### 4.2 User Interface

Fig. 1 shows an overview of the Data Therapist interface. The view is dominated by a scrollable spreedsheet view of the current dataset. This interface is shown as soon as the user uploads a dataset into the tool.

The **basic Data Therapist workflow** involves the user continuously choosing between three options:

(1) Annotate the overall dataset (General Annotation);
(2) Annotate a specific region of the dataset (Data-Specific Annotation); or
(3) Answer a Data Therapist-initiated question about the dataset (Question-Answering).

Below we detail these three mechanisms (Figure 2): (1) general annotation, (2) data-specific annotation, and (3) question-answering.

### 4.3 General Annotation

Data Therapist allows the user to add general annotations to the dataset by entering text into the text box (see Figure 1f) below the main spreadsheet view. The implication is that these annotations apply to the full dataset and **not** to specific regions (cells, columns, or rows). There is a wide range of information that the user can provide about the dataset itself. For example, one can specify where the dataset originated, why it is important, its intended purpose, potential applications, and its broader significance. Each time an annotation is uploaded to the interface, the annotation is stored in the server, and is shown in a list in the annotation view (see Figure 1B) (DC5).

### 4.4 Data-Specific Annotation

Annotating individual data entries allows the user to provide fine-grained context or observations about specific portions of the dataset. There are several methods to create this type of annotations (DC2).

First, they can directly select from the spreadsheet (Figure 1C). The spreadsheet enables selecting multiple instances by row, column, or individuals cells, as well as using a rectangular rubberband selection. It also allows reordering the dataset as well as changing the sorting order.

Second, users can choose instances from visualizations automatically created by selecting columns. The visualizations helps users quickly understand and read patterns in the dataset that would be difficult to see in the table view (DC3). This helps the user not only to understand the data, but also easily select the instances that they want to annotate. Our implementation currently supports creating histograms and scatterplots. By right-clicking on a column header in the spreadsheet, the user can choose from several icons. Selecting the first icon sorts the column's values according to its feature. Selecting the second icon displays a histogram depicting the column's distribution next to the spreadsheet (see Figure 1e). The next two icons allow the user to designate two axes for a scatterplot. When multiple visualizations are shown in the interface, these visualizations work as a cross-filtering functions, where only corresponding instances in the visualizations are selected. The user can add annotations to the selected location using the textbox in Figure 1f.

In addition, to help users locate the relevant data, hovering over a data-specific annotation in the annotation view (Figure 1B) highlights the corresponding instance(s) in the spreadsheet (Figure 1c).

### 4.5 Question-Answering

Finally, the last approach is to elicit domain knowledge from the user by having them answer the questions automatically generated by Data Therapist in the *question view* (Figure 1A). Question and answering works in the following manner (question generation is is explained in § 4.6):

(1) Data Therapist initializes the question view list with 10 questions relevant to that dataset.
(2) The user can choose to answer any question in the list by clicking on it, opening up a textbox where they can write their response.
(3) If a question is irrelevant, the user can remove it from the list.
(4) The user may refill the list of questions in the question view by clicking "generate new questions."

To support the question-answering process, Data Therapist provides two key features: (1) verification of user responses, and (2) automated thematic summarization of predefined questions. After the user submits an answer, the system verifies whether the response adequately addresses the question. If the answer passes this check, the question disappears from the list after a brief delay. Otherwise, the system displays a message explaining why the answer was rejected, allowing the user to revise and resubmit.

Specifically, the system evaluates two criteria: (1) whether the answer faithfully addresses the question, and (2) whether it conflicts with any existing annotations. This process is powered by ChatGPT, as detailed in § 4.7. We illustrate how a response is validated in Figure 3. Also, once the user has answered enough predefined questions (from when the user answered 2 questions per theme), Data Therapist automatically generates a thematic overview of those responses. This summary clarifies which metadata fields have been completed and which remain unfilled. Finally, to facilitate sharing the accumulated information, the "export annotations" feature (Figure 1g) enables users to download a .JSON file containing all annotations.

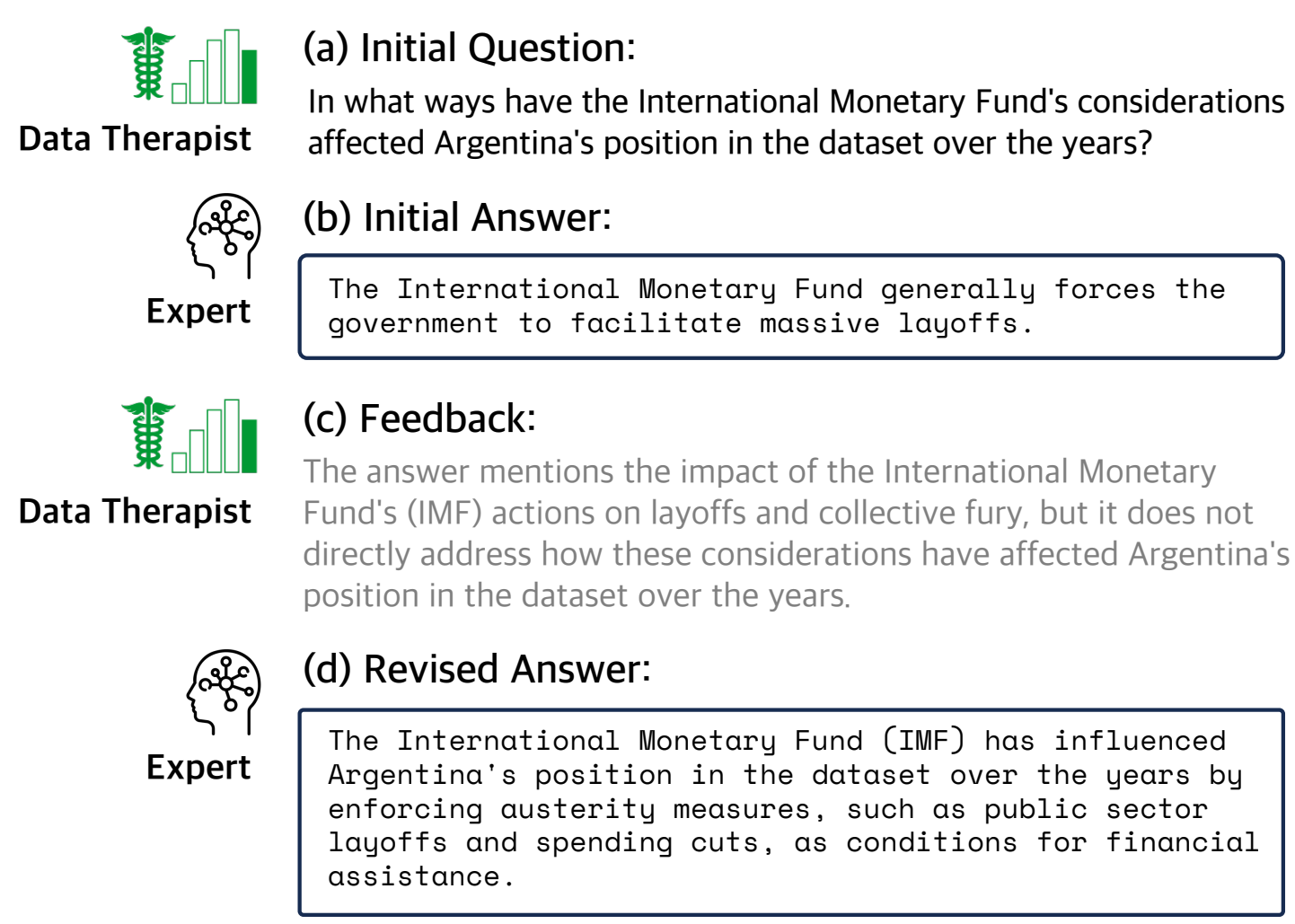

Fig. 3. **How Data Therapist validates responses.** Given an initial question (a), the expert provides an answer (b). The Data Therapist system then evaluates it, either validating or rejecting it with feedback (c). If rejected, the expert can submit a revised answer until it is validated (d).

### 4.6 Generating Questions

Here we describe the mechanism behind how the questions are generated and presented to the user.

*Generating questions.* In the backend, as soon as the dataset is uploaded, the LLM generates 30 questions that aim to elicit domain knowledge and information from the user. We maintain all questions in a single list and use a custom prioritizing function to select those that best align with our priorities whenever a request is made. When a dataset is first created, we generate 30 questions for two reasons. The first is to provide an initial set of questions as a starting point for the annotation process. The second is to prevent running out of questions in cases where no annotations are made but the user asks for more questions, since new questions are only generated upon user annotation. Each time the user leaves an annotation, five additional questions related to that annotation are generated. If there are fewer than ten LLM-generated questions left in the list, we immediately add ten more to maintain a sufficient number of available questions. Below we describe the prioritizing function used to select the questions.

*Prioritizing questions.* The prioritization of questions are partially motivated by the methodology conducted by Park et al. [37]. As shown in Equation 1, the score of a question is based on three factors: originality, recency, and importance. We let each metric be integers between 1 and 5 in Likert scale. We add scores from these three factors, and select from the highest-scoring questions.

$$\text{Score}_q = \text{Originality}_q + \text{Recency}_q + \text{Importance}_q. \tag{1}$$

The originality of a question, $\text{Originality}_q$, measures how distinct a given question is relative to existing annotations. We calculate it by first computing a similarity score based on the arithmetic average between the question and previously used annotation questions, then scale this score from 1 to 5 (rounded to integers). The recency, $\text{Recency}_q$, tracks how recent the question is. Whenever a new annotation is created, the question starts with a recency score of 5. Each time the user adds an annotation, this score decreases by 1, but never below 1. Finally, importance, $\text{Importance}_q$, indicates how critical the question is to our primary goal of understanding the dataset. We instruct an LLM to evaluate (1) the question's ability to clarify the dataset and (2) its potential to introduce novel insights.

In addition, when prioritizing predefined questions, we prioritize those from the theme with the most unanswered items. This approach ensures an even update of our thematic summary. If two questions share the same priority level, one is selected at random.

*Filling in questions.* The initial question view is populated with 10 questions. Among them, half of them are from the list of predefined questions by Gebru et al. [12], and the other half are selected from the list of LLM-generated questions. We recommend predefined questions for two main reasons: first, these datasets were originally created to describe themselves; second, we seek to further enrich their metadata. When filling in questions, the interface displays up to 10 questions at a time. A question can be removed in one of two ways: either by answering it or by manually deleting it. The "generate new questions" button becomes active once the interface has at most four remaining questions. At that point, half of the empty slots are populated with predefined questions, and the other half are populated with LLM-generated questions. If the number of empty slots is odd, one additional LLM-generated question is added. For example, if the user presses the "generate new questions" button when only two questions remain, there are eight empty slots to fill, resulting in four predefined questions and four LLM-generated questions. However, if just one question remains, there are nine empty slots to fill, which will be split into four predefined questions and five LLM-generated questions. Additionally, once all predefined questions are exhausted, the system provides only LLM-generated questions.

Table 1. **Study demographics.** A total of 6 domain experts from 3 different expertise (Accounting, Political Science, and Usable Security) participated in our user study. The participants were all above the age 18. For each domain, we tried our best to bring experts with similar years of experience in the field. If this was not feasible, we ensured that each expert had at least three years of professional experience in their field.

| Topic | ID | Gender | Age | Education | Job Title | Experience | Assistance from |
|---|---|---|---|---|---|---|---|
| Accounting | A1 | Male | 33 | B.S. | Certified Public Accountant | 7 yrs | Data Therapist |
| | A2 | Male | 34 | B.S. | Certified Public Accountant | 6 yrs | Human interviewer |
| Political Science | P1 | Male | 37 | Ph.D. | Assistant Professor in Political Science | 8 yrs | Data Therapist |
| | P2 | Male | 36 | Ph.D. | Research Professor in Political Science | 8 yrs | Human interviewer |
| Usable Security | S1 | Male | 34 | Ph.D. | Ph.D. Student in Usable Security | 2.5 yrs | Data Therapist |
| | S2 | Female | 30 | B.S. | M.S. Student in Usable Security | 2.5 yrs | Human interviewer |

### 4.7 Prompt Engineering

The Data Therapist takes a human-in-the-loop approach, where human knowledge is added to the system, and the system uses it to create questions generated from LLMs, and again, the questions help users elicit more detailed explanations and knowledge. We structured this procedure by employing the following 5 strategies. To this end, we refer to both prior work and instructions provided in the ChatGPT API [36].

- **Describe the role and task precisely** – to ensure responses faithfully align with our intent.
- **Structure responses clearly and concisely** – to maintain precision and avoid verbose outputs.
- **Guide a specific response template** (e.g., using delimiters for outputs) – to foster consistency and streamline parsing.
- **Detail context and provide all relevant information** – to anchor the system's output in accurate domain knowledge.
- **Use global and local variables in the prompt** – to maintain consistent references across global (e.g., specifying the role of Data Therapist) and local (e.g., tasks) specifications.

Below we detail the prompt templates used in the Data Therapist. We first show how we specify the role of Data Therapist in the interaction loop. Next, we present templates used for generating and validating questions. We lastly describe templates for smaller tasks, such as summarizing texts and assessing the importance of questions. The full prompt details are provided in the OSF submission [1].

*Conceptualizing Data Therapist.* In defining the role of a "data therapist," we sought to describe it in detail, clarifying its objectives, the problems it addresses, and the responsibilities we expect it to fulfill. This prompt serves as a global variable for each task prompt. Defining this as $[\text{Role}]_{DT}$ is as follows:

> $[\text{Role}]_{DT}$: *The role of the Data Therapist is to elicit knowledge about the dataset from users by asking appropriate questions that, by answering, could help them understand their own data. The questions should aim to bridge the gap between reality (the situation, environment, and background surrounding the dataset) and the dataset itself. While the data is present, it does not, on its own, explain the background or issues. Ideally, annotations should be made so that even someone unfamiliar with the dataset can understand it by reading those annotations. Although the main goal is to help extract annotations by asking questions, the Data Therapist can also assist in other tasks related to annotation, such as validating questions.*

[1] https://osf.io/e4fxu/?view_only=0ac12e1c4b8e4b99a5739935aa0ca97e

*Generating questions.* We generate two types of questions. The first type focuses on the dataset itself, while the second type is derived from the annotations. In generating $T1$, we leverage ChatGPT's built-in RAG and in-context learning (ICL) functionalities, while for follow-up questions in $T2$ we embed the user's previous, relevant responses into each subsequent query to the LLM. Below, we present the prompt templates used for generating each type of question.

We define $\mathbf{T}_{T1}$ as the template for generating the first type of question. Let $[\text{Role}]_{DT}$, and $[\text{Output-format}]_{T1}$ be textual strings, respectively, that prompts the role of the Data Therapist, the task of generating this question type, and the specific output format (for $\mathbf{T}_{T1}$, we present a json-style example of an output). {Dataset} is the dataset of interest in a textual form. We define $[\text{Task}]_{T1}$ as follows:

> $[\text{Task}]_{T1}$: *Read the dataset. Generate 30 key questions you would ask the experts of the dataset that would help people who do not know about the data understand the dataset. The generated questions should (1) help clarify the dataset from a non-expert's perspective and (2) bridge the gap between the explanations and the dataset itself.*

Then, summing them all up, we define $\mathbf{T}_{T1}$ as:

$$\mathbf{T}_{T1} = [\text{Role}]_{DT} + \{\text{Dataset}\} + [\text{Task}]_{T1} + [\text{Output-format}]_{T1}.$$

$\mathbf{T}_{T2}$ refers to the template for generating the second type of question. We define $\{\text{Questions}\}_{all}$ and $\{\text{Questions}\}_{rec}$ as, respectively, the full set of questions used for annotation and the most recently annotated question (if available). $\{\text{Annotations}\}_{all}$ and $\{\text{Annotations}\}_{rec}$ are the complete set of annotations and the most recent one, respectively. Let $[\text{Output-format}]_{T2}$ and $[\text{Task}]_{T2}$ be a JSON-style output format for this question type and the textual description of the corresponding task. We define $\mathbf{T}_{T2}$ as follows:

$$\begin{aligned}\mathbf{T}_{T2} = [\text{Role}]_{DT} + \\ &\{\text{Dataset}\} + \{\text{Questions}\}_{all} + \{\text{Annotations}\}_{all} + \\ &\{\text{Questions}\}_{rec} + \{\text{Annotations}\}_{rec} + [\text{Task}]_{T2} + \\ &[\text{Output-format}]_{T2}.\end{aligned}$$

*Validating answers.* There are two types of validations for the answers checking (1) if the answer correctly addresses the question, and (2) for contradictions across questions and answers. The template for the first type of validation ensures that the answer aligns with the intent of question. For this, we define $[\text{Task}]_{T1}$ as:

> $[\text{Task}]_{V1}$: *Now, look at the answer and check if the answer makes sense, based on the question. If the answer makes no sense at all, then provide feedback to guide users to answer the question.*

Then the $\mathbf{T}_{V1}$ becomes:

$$\begin{aligned}\mathbf{T}_{V1} = [\text{Role}]_{DT} + [\text{Question}]_{test} + [\text{Answer}]_{test} + \\ &[\text{Task}]_{T1} + [\text{Output-format}]_{V1}.\end{aligned}$$

The template for the second type of validation ($\mathbf{T}_{V2}$) is constructed in a similar manner, comparing existing annotations with the most recently updated one. Additionally, we implement auxiliary text processing functions, such as text summarization. All prompt templates and further explanations are provided in our supplementary materials.

### 4.8 Implementation

We implement Data Therapist as a client-server web application. The front-end of the system is implemented using `JavaScript` and `React` framework, where it communicates with the `Python Flask` server using `axios`. The backend server was hosted on Amazon Web Services (AWS). As the LLM, we use the API provided by OpenAI. In particular, we use `o1` for initial question generation to leverage its reasoning capability, and `GPT-4o` for all other LLM calls.

Table 2. **Location-tagged annotations.** A large part of annotations created are location-tagged (in spreadsheets). This pattern is more salient in human-guided annotations.

| | Accounting | | Political Science | | Usable Security | |
|---|---|---|---|---|---|---|
| | DT | HI | DT | HI | DT | HI |
| Without annotations | 11 | 2 | 10 | 0 | 13 | 0 |
| Location-tagged annotations | 13 | 22 | 25 | 35 | 10 | 15 |

Table 3. **Annotation patterns.** Below, we present annotation statistics from the four domain groups that participated in our study: Accounting, Political Science (Political Sci.), and Usable Security. The second row shows the number of annotations generated by participants who collaborated with a human interviewer, including the total number of annotations (Total annotations) and a breakdown of those created with and without interviewer assistance (Anno. w/ human inter., Anno. w/o human inter.). The third row shows annotations from participants who used Data Therapist, including the total number, those created without the system (Anno. w/o Data Therapist), and those generated in response to LLM-generated questions (Ans. from q. by LLMs) or predefined questions (Ans. from q. by predef.). The bars and percentages in parentheses indicate each category's proportion relative to the total annotations per participant.

| | Metric | Accounting | Political Science | Usable Security | Total |
|---|---|---|---|---|---|
| **w/ human interviewer** | Total annotations | 24 | 35 | 15 | 96 |
| | Anno. w/o human inter. | 12 (50%) | 12 (34%) | 10 (67%) | 46 (48%) |
| | Anno. w/ human inter. | 12 (50%) | 23 (66%) | 5 (33%) | 50 (52%) |
| **w/ Data Therapist** | Total annotations | 24 | 32 | 24 | 108 |
| | Anno. w/o DT | 7 (29%) | 17 (53%) | 9 (38%) | 40 (37%) |
| | Ans. from q. by LLMs | 15 (63%) | 10 (31%) | 10 (42%) | 47 (44%) |
| | Ans from q. by predef. | 2 (8%) | 5 (16%) | 5 (21%) | 22 (19%) |

Table 4. **Difficulty of annotations.** We ask the domain experts in Accounting (Acct.), Political Science (Pol. Sci.), and Usable Security (Usab. Sec.) to evaluate their own annotations as well as those from the other expert with different experiment conditions. We let them assess from two perspectives: informativeness (the degree of which the annotation helps explain the data), and difficulty (the degree to which the annotation is difficult). The detailed evaluation criteria are described in § **??**.

| | | **Difficulty** | | | |
|---|---|---|---|---|---|
| | | **Acct.** | **Pol. Sci.** | **Usab. Sec.** | **Mean** |
| **w/ human interviewer** | Anno. w/o human inter. | 2.75 | 2.18 | 1.90 | 2.43 |
| | Anno. w/ human inter. | 2.59 | 2.25 | 1.90 | 2.34 |
| **w/ Data Therapist** | Anno. w/o DT | 2.07 | 2.42 | 1.78 | 2.57 |
| | Ans. from q. by LLMs | 2.10 | 2.15 | 1.65 | 2.22 |
| | Ans. from q. by predef. | 1.00 | 1.80 | 1.60 | 1.85 |

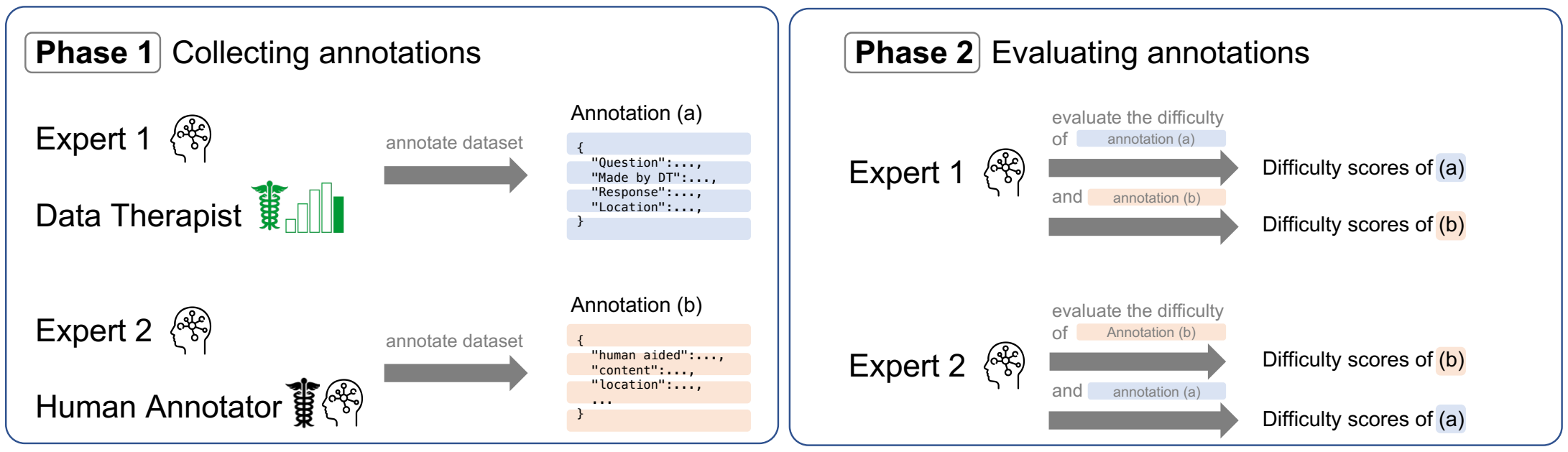

Fig. 4. **Experiment setup and evaluation.** In each dataset scenario (we have three scenarios: Accounting, Political Science, and Usable Security). Two domain experts are involved in each scenario. In phase 1, one expert uses DT to make annotations, and another expert uses human annotator to make annotations. Then, in phase 2, each expert evaluate the difficulty score of annotations made by them and the other expert, from 1 to 5 scale (the scale is explained in § 5.3.

## 5 Evaluation Setup

We conduct our user study to examine how the Data Therapist feedback system differs from human annotation. Our methodology is based on using a human interviewer's annotations as a baseline [8, 22, 54]. We adopt this approach for two reasons. First, it lets us understand how closely an LLM-driven mechanism can match human performance. Second, benchmarking Data Therapist directly against human annotators—and understanding its gaps—provides us a direction for future improvements.

Our evaluation is structured as follows: using a single domain-specific dataset, we elicit annotations from two sources, Data Therapist and a human assistant (two experts on the same topic). We treat the human interviewer's domain knowledge as the gold standard and assess how closely Data Therapist aligns with it. The details about the task is described in § 5.2. The study was organized into two sessions, each around 2 hours, comprising three parts in total. Our experiment was approved by the national institutional review board of South Korea. Below we provide details of the user study.

### 5.1 Participants

We recruited a pair of experts for four different domains, yielding a total of 8 participants. We screened participants to have at least an undergraduate degree, at least two years of research or practical experience in the field, fluency in reading, writing, and speaking English, and, most importantly, comparable career experience for both experts in a given domain. By applying these criteria, we selected pairs of participants in three domains: Accounting, Political Science, and Usable Security. Each participant received a $20 gift card as compensation upon completing the study. The demographics of the participants are shown in Table 1.

### 5.2 Task

We compared the effectiveness of the Data Therapist system with that of a human interviewer. For each pair of domain experts, our experiment asked one expert (HI) to collaborate with a human interviewer (the study administrator), while the other expert (DT) exclusively used the Data Therapist system. DT participants were free to add annotations at their discretion and could also decide whether or not to seek help or respond to questions generated by Data Therapist. Each participant was given approximately 45 minutes to annotate the dataset.

Prior to the experiment, we asked each pair of domain experts to together select a type of data, while familiar to both, would be difficult for non-experts to understand. We asked for a small, manageable example tabular dataset (not more than 10,000 rows and 20 columns). After a brief email exchange, they chose the following datasets.

- **Accounting:** A cash flow statement for a resort company.
- **Political Science:** The Political Terror Scale dataset.[2]
- **Usable Security:** Computer Network Traffic dataset.[3]

### 5.3 Procedure

We conducted the study entirely online through video teleconferencing in individual sessions. The study consisted of two sessions with three phases: (1) preparation and instruction, (2) annotation, and (3) domain knowledge assessment. Participants completed Phases 1 and 2 in the first session and Phase 3 in the second session.

*Phase 1: Preparation and Instruction.* The instructor first collected each participant's information via a consent form. After starting each session, the instructor outlined the research problem and objectives, clarifying expectations for each participant. For DT participants, we explained all of Data Therapist's features and ensured that they could use all tools effectively. For HI participants, we presented the dataset in a Google Spreadsheet and asked the participant to enter annotations directly there. We also requested that any annotations generated from working with the human assistant include '[H]' at the end. This phase took approximately 15 to 20 minutes.

*Phase 2: Annotation.* All participants were given 45 minutes to annotate and describe the dataset. DT participants began by creating annotations independently. If they encountered difficulties or saw a question they wished to answer, they did so and reviewed any answered questions. They were also able to generate visualizations, especially for large datasets (e.g., in usable security) to identify patterns.

HI participants started by adding annotations they believed were important. We asked them to explicitly write responses and annotations using their keyboard rather than verbally—which might be more natural and similar to a typical domain expert interview—to ensure parity with the DT participants. If they became stuck or needed help, or if no activity occurred for over 20 seconds, the human interviewer provided feedback on missing or insufficient explanations. Participants could accept this feedback and update their annotations or ignore it if they deemed it unnecessary. In both conditions, participants were notified when there were 10 minutes and 5 minutes remaining.

*Phase 3: Design knowledge assessment.* We conducted the second session at least a day after the first session ended. Here participants evaluated both their own annotations and those made by others, whether through Data Therapist or human assistance, using a 1 to 5 scale. We asked participants to assess based on the annotation's **difficulty**. Of course, difficulty is not a proxy for tacit knowledge. However, high difficulty level suggests the dataset includes concepts that are hard for non-experts, or visualization designers, to grasp. Our goal here is to assess if LLMs can ask difficult questions like humans do. For difficulty, a rating of 1 denoted 'comprehensible to a lay audience,' 3 indicated 'known to all domain experts but not to lay audiences,' and 5 signified 'understandable only to certain specialized domain experts.'

[2]https://www.politicalterrorscale.org/Data/Download.html
[3]https://www.kaggle.com/datasets/crawford/computer-network-traffic/data

### 5.4 Apparatus

The interview and experiment were conducted using the participants' own computers. To make the environmental setting consistent and convenient, we recommended the participants to use Google Chrome, and zoom out to 70% in a $1920 \times 1080$ resolution display.

### 5.5 Data Collection

We gathered participant demographics via an online form. With their permission, we recorded and filmed each participant's experimental session. We also compiled information on the questions they used and the annotations they added to the dataset. In addition, during the second session, we gathered the scores provided by participants in their evaluations.

## 6 Evaluation Results

We now evaluate Data Therapist. Because our design choices emphasize structured communication between experts and systems, we first assess how well expert interactions align with our design intentions (§5.4). Next, we evaluate the effectiveness of Data Therapist by comparing its annotations with those produced by human experts (§6.2). We then analyze cases in-depth by measuring the difficulty of annotations to further understand when LLM-driven questions, or experts perform particularly well, and why (§6.3).

### 6.1 Expert Interaction with the Data Therapist

In particular, we are interested to see how our tools support multi-granular annotation and analysis (DC2, DC3), and facilitate comprehension coverage (DC6). Tables 2 and 3 show the annotation patterns while the experts made annotations.

To begin with, our tool helps expand coverage (DC6) by making experts answer questions generated in our system. As a case in point, DT participants on average created 37% of their annotations without help from the tool, but this figure fluctuated considerably per participant. In particular, LLM-generated questions contributed to creating a substantial number of questions (DC1). Table 3 shows that 31 to 63% of questions in DT are answers from questions generated by LLMs. Although the system initially suggested an equal mix of LLM-generated and predefined questions, approximately 60% of the questions in DT were produced by the LLM, while the remaining 40% were drawn from the predefined questions by Gebru et al. [12]. Moreover, as a sidenote the absence of direct human contact slightly increased the number of annotations. On average, DT participants produced four more annotations per group than HI participants.

Furthermore, we identify that supporting multi-granular annotations, as well as enabling analysis of the data is an important design point. Table 2 shows that a large part of annotations pertain to parts of the spreadsheet either selected (DC2) by the expert, or filtered using cross-filter functions (DC3).

### 6.2 Effectiveness of the Data Therapist

We next evaluate the effectiveness of the Data Therapist in eliciting tacit knowledge.

*Methodology.* We compare the annotations derived from human- and Data Therapist-generated questions to examine how Data Therapist bridges reality and representation. Specifically, we analyze the topical and functional similarity of annotations produced by DT and HI. Because exact matches were rare, we treated annotations as similar when their topics and goals aligned, even if phrased slightly differently. For example, in the PTS dataset, the annotations "*Starting* *from level 2, there is some political violence ...*" and "*Level 1 means there is almost no government-led violence ...*," were grouped together because both define the intensity of political terror-scale. Details of our grouping criteria are provided in our OSF repository. Based on this process, we classify topics into three categories: (1) shared—appearing in both DT and HI, (2) DT-exclusive, and (3) HI-exclusive.

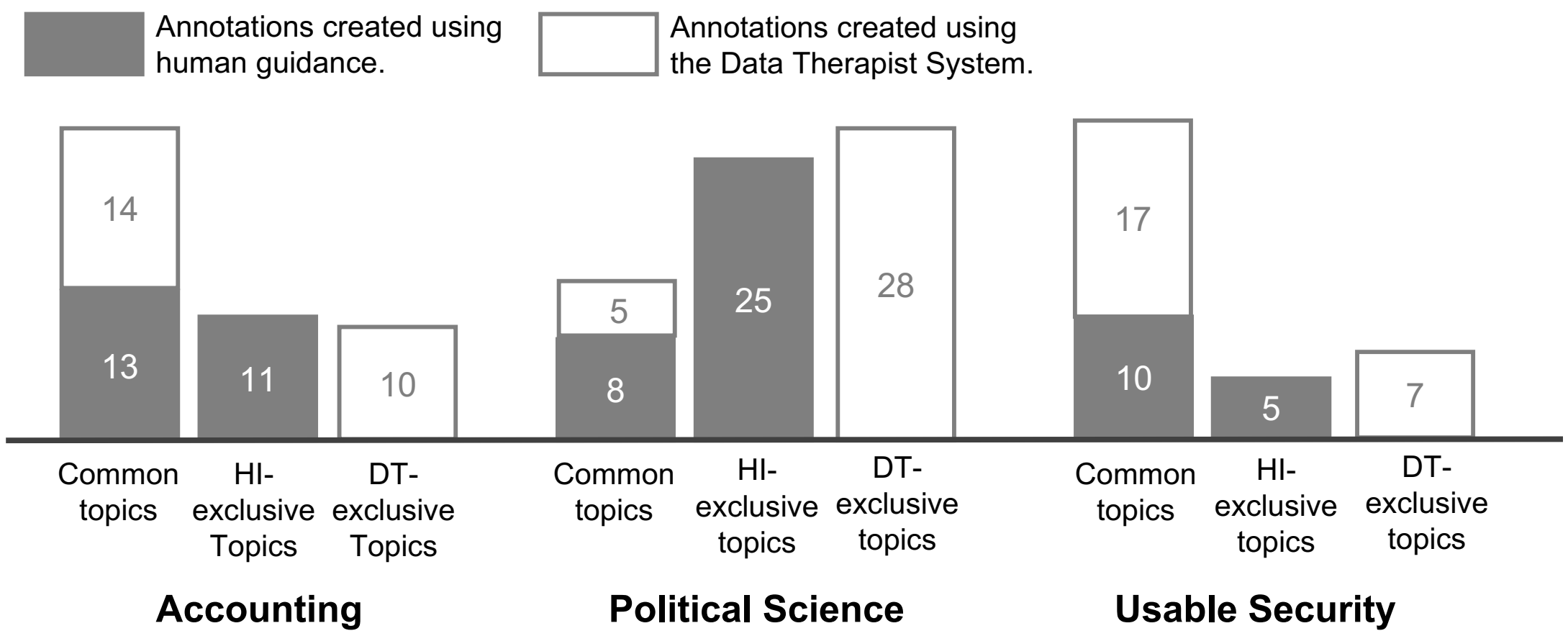

Fig. 5. **Annotations generated with Data Therapist vs. human guidance.** Annotations are classified into 3 categories: shared topics, human(HI)-exclusive topics, and Data Therapist(DT)-exclusive ones. Gray-filled bars indicate annotations generated with DT, while gray-outlined bars indicate annotations generated through human guidance. The greater the number of common topics, the more effective DT is. HI-exclusive or DT-exclusive topics indicate cases where humans or DT are uniquely effective.

*Results.* Figure 5 summarizes the number of annotations from DT and HI participants, classified as Common, DT-exclusive, and HI-exclusive topics across four areas of expertise. Filled bar represents annotations from DT, and stroked bar represents those from HI. The greater the number of common topics, the more effective DT is, while HI-exclusive or DT-exclusive topics count cases where humans or DT are uniquely effective. Although the distribution of shared and exclusive topics varies by dataset, several recurring patterns emerge across areas, which we describe below.

*We first found that Data Therapist operates as designed.* We observe that in Accounting and Usable Security datasets, we identify that more than a majority of DT annotations (58% and 71% ) are similar to annotations created in HI. By analyzing common topics between DT and HI, we identified two recurring patterns: (1) annotations that explain how to read or interpret the dataset (e.g., *"Number 88 means data unavailable."* in Political Science). (2) those that clarify difficult vocabularies or variables to lay audiences (e.g., *"A Remote Autonomous System Number (Remote ASN) is a numerical identifier for an Internet Service Provider."* in Usable Security). For example, in Accounting, both DT and HI focused on understanding the cash flow mechanism in financial statement, while in Usable Security, the common topic was on methodologies for detecting anomalies in network traffic.

However, we also identify differences between HI- and DT-exclusive topics, as 42% and 29% of DT annotations do not align with HI annotations. One reason is due to the difficulty in predicting what the lay audience may not know. While annotator in HI has a clear idea of what the they do or do not know, the LLM's knowledge level of a lay audience may not equate to that of a human annotator. Another reason stems from making annotations from pre-defined questions [12]. Note that not all questions from DT are created using LLMs, and pre-defined questions are suggested in Data Therapist. We find that none of the pre-defined questions, although chosen by the annotator align with topics in HI.

In contrast, in the PTS dataset (Political Science), we identify that only 15% of annotations in DT (5/28) overlap with those in HI. This low level of disaccordance stems from both the intrinsic nature of the dataset and the differing expertise of the two Ph.D annotators. The common topics centered on the definition of key variables specific to each dataset. However, since the dataset covers global information, both experts have made a number of annotations by selecting different areas of the spreadsheet to highlight local context that explain PTS values of their interests. Although both hold PhDs in Political Science, their regional specializations differed—one in Southeast Asia and the other in Northeast Asia. This contributed to the difference in our results.

### 6.3 In-Depth Analysis of our Results.

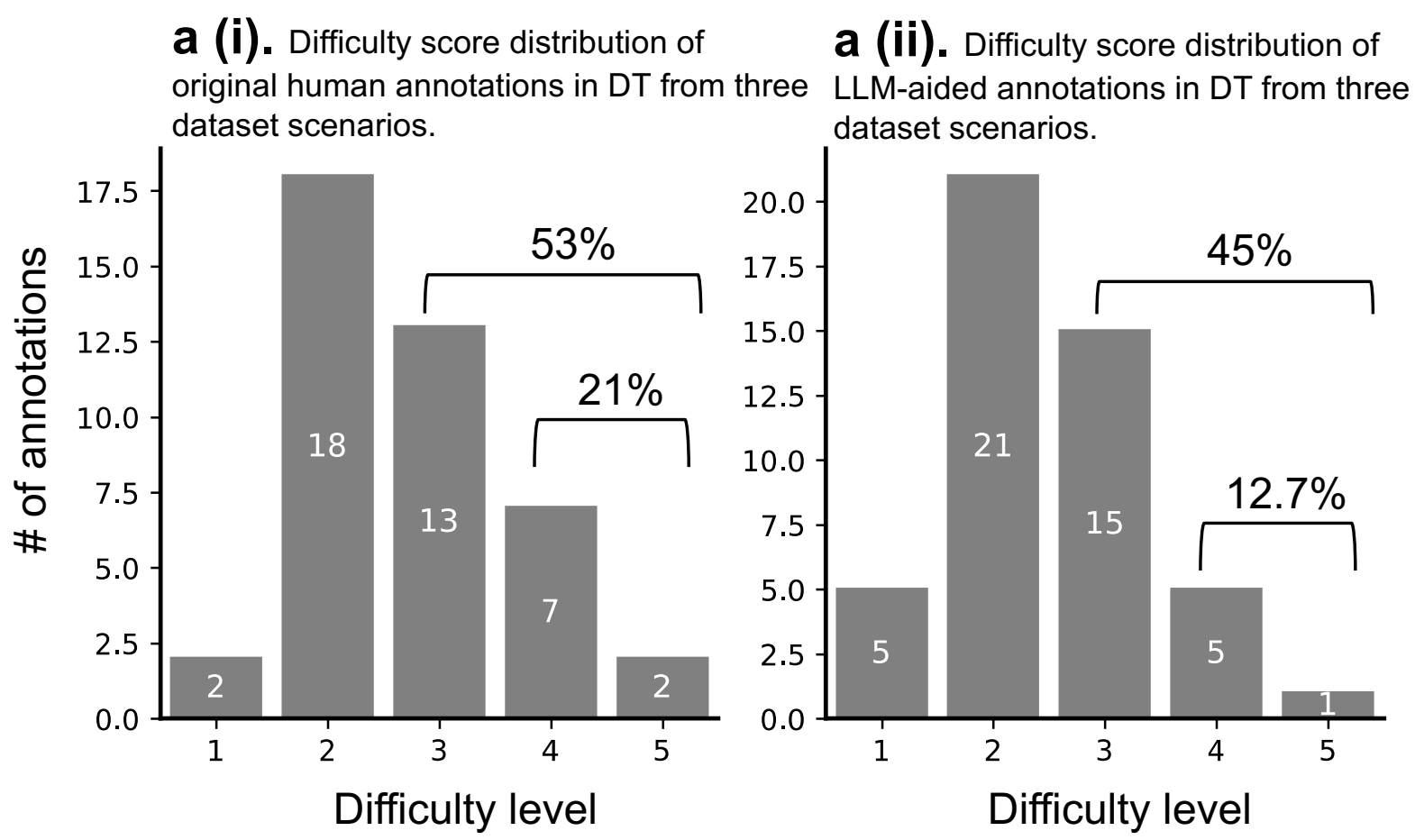

Fig. 6. **Analyzing the difficulty of annotations.** To understand the impact of difficulty in annotations, we make experts evaluate the difficulty of each annotation. From this, we find out that while LLMs can pose difficult questions, their ability is not yet to the human standard.

Difficulty is not a proxy for tacit knowledge, but being able to ask difficult questions is important, because they are hard for non-experts, or visualization designers, to grasp. Hence, understanding difficult knowledge should not be overlooked. Figure 6 shows two charts that show the distributions of annotations per different difficulty levels. The left (i) shows the original human annotations, and the right (ii) shows annotations created using LLMs. In general, the distribution trend of difficulty levels in human and LLM-aided annotations seem to be similar, where most annotations lie between 2 and 3. However, we identify several differences in low and high scores. First, LLMs can sometimes ask 1-level questions that everyone seem to know. While it could be redundant to most people, such inquiries can help lay audience who really do not know such questions. In a way, it adheres more towards lay audience than what experts think. This is an advantage of LLMs that experts do not possess, considering that the evaluation is done by the experts.

However, the story becomes different when looking for difficult questions. Although humans are 8% points better than LLMs, 45% of annotations higher or equal to 3 means that LLMs are capable of providing questions on 3-difficulty level. However, compared with human annotations (21%), the percentage of annotations from 4 or over-difficulty level reduces to 12.7%. These results suggest LLMs can pose tough questions but are not yet at the human standard.

The analysis of difficult questions that LLMs can and cannot yet answer provides us where the true challenge lies. First, we find that LLMs can guess difficult questions that require interpretations or understand implications about a topic, such as understanding cash flow mechanisms. However, it has two observable problems in that it is having problem connecting the unfamiliar, but latent event, that can only be acquired through expert knowledge (e.g., *'In Myanmar, in Feb 2021, the military junta carried out a successful coup and toppled the then democratic regime.'* (Political Science) ).

## 7 Discussion

Below we discuss the implications and limitations behind Data Therapist.

### 7.1 What Part of Data Therapist Is Effective and What Part Isn't.

In our study, we compare the annotations generated through human assistance with those produced by LLMs via question-based prompts. One of the objectives in the experiment was to use LLMs and identify if LLMs could function like humans in help generating annotations.

We find that the Data Therapist system provides various functionalities to help support multi-granular annotation and facilitate analysis. The results show that a significant number of annotations are location-tagged, which signals that our design was effective in supporting experts' annotations. However, one limitation we have is that we tested using relatively small dataset, and the issue remains to find effective methods to deal with large-scale spreadsheets. We aim to challenge this issue as our future work.

Next, LLMs can generate questions in a manner similar to humans, particularly when it comes to understanding the underlying mechanisms within a dataset and guessing knowledge that lay audience may not know. We can also identify that the questions from LLMs definitely align more with humans than the list of predefined questions. This has been proved through our studies. One possible explanation for this strength is that LLMs may have previously encountered similar datasets used by our participants, or have prior knowledge of the data. In fact, several participants made similar comments. However, because in reality, lay audience are not perfectly identical, LLMs can sometimes miss out on suggesting or explaining definitions. As our future work, we aim to develop a new mental model, using LLMs to simulate how a lay audience would react on a new topic, or dataset.

While Data Therapist is clearly no replacement for a human interviewer, we suggest that Data Therapist be used as a simulacrum [38] to support the initial stages of data collection and screening to bridge the gap between real-world context and the data. In this role, it can serve as a starting point for conducting broad, general surveys, enabling human experts to concentrate more effectively on domain-specific details.

### 7.2 Difficulty of Assessing Annotations.

It is difficult to completely fairly evaluate the annotations. We argue that it is difficult to recruit two experts of the same level of knowledge base and skills. We have identified from our experiment that annotations can result very differently upon the different background of the annotator, or the expert. A good example of this can be seen from our Political Science case, where the experts had different knowledge base. These experts had the same years of experience, same Ph.D. title, similar age and gender, but their knowledge base created a huge difference in annotation patterns.

There are also various factors that can further influence the results. For example, another factor that influences the quality of questions is the seniority level of the experts. We computed the correlation between the annotations created by different years of experience of the expert. We can see that in both cases the correlation line tends to be positive in both DT (0.917) and HI (0.713). Such level could have had impact on the results of our work.

As a future work, we aim to find additional biases that could impact our design experiment methodologies, and try to be more rigorous in conducting the experiment.

## 8 Conclusion

We presented Data Therapist, a structured mixed-initiative interaction approach to question-answering and user-driven annotation for eliciting domain knowledge about a dataset. In doing so, we developed the Data Therapist implementation as a practical web-based tool. We conducted a user study involving four groups, each consisting of two domain experts with similar expertise. Our findings indicate that while LLM-generated questions are effective in helping users understand the complex mechanisms embedded in the dataset, accurately estimating the knowledge level of a lay audience remains an interesting avenue for future work.

## Acknowledgments

This work was supported partly by Villum Investigator grant VL-54492 by Villum Fonden. Any opinions, findings, and conclusions expressed in this material are those of the authors and do not necessarily reflect the views of the funding agency.

## References

[1] Rohan Anil, Andrew M. Dai, Orhan Firat, Melvin Johnson, Dmitry Lepikhin, Alexandre Passos, Siamak Shakeri, Emanuel Taropa, Paige Bailey, Zhifeng Chen, Eric Chu, Jonathan H. Clark, Laurent El Shafey, Yanping Huang, Kathy Meier-Hellstern, Gaurav Mishra, Erica Moreira, Mark Omernick, Kevin Robinson, Sebastian Ruder, Yi Tay, Kefan Xiao, Yuanzhong Xu, and Yujing Zhang et al. 2023. PaLM 2 Technical Report. *CoRR* abs/2305.10403 (2023), 93 pages. arXiv:2305.10403 doi:10.48550/arXiv.2305.10403

[2] Alex Bigelow, Steven Drucker, Danyel Fisher, and Miriah Meyer. 2014. Reflections on how designers design with data. In *Proceedings of the ACM Conference on Advanced Visual Interfaces*. ACM, New York, NY, USA, 17–24. doi:10.1145/2598153.2598175

[3] Tom Brown, Benjamin Mann, Nick Ryder, Melanie Subbiah, Jared D. Kaplan, Prafulla Dhariwal, Arvind Neelakantan, Pranav Shyam, Girish Sastry, Amanda Askell, et al. 2020. Language models are few-shot learners. In *Advances in Neural Information Processing Systems*, Vol. 33. Curran Associates, Inc., Red Hook, NY, USA, 1877–1901.

[4] Alyxander Burns, Christiana Lee, Thai On, Cindy Xiong, Evan Peck, and Narges Mahyar. 2024. From Invisible to Visible: Impacts of Metadata in Communicative Data Visualization. *IEEE Transactions on Visualization and Computer Graphics* 30, 7 (2024), 3427–3443. doi:10.1109/TVCG.2022.3231716

[5] Alyxander Burns, Thai On, Christiana Lee, Rachel Shapiro, Cindy Xiong, and Narges Mahyar. 2021. Making the Invisible Visible: Risks and Benefits of Disclosing Metadata in Visualization. In *Proceedings of the IEEE Workshop on Visualization for Social Good*. IEEE Computer Society, Los Alamitos, CA, USA, 11–15. doi:10.1109/VIS4Good54225.2021.00008

[6] Qing Chen, Fuling Sun, Xinyue Xu, Zui Chen, Jiazhe Wang, and Nan Cao. 2022. VizLinter: A Linter and Fixer Framework for Data Visualization. *IEEE Transactions on Visualization and Computer Graphics* 28, 1 (2022), 206–216. doi:10.1109/TVCG.2021.3114804

[7] Wei-Lin Chiang, Zhuohan Li, Zi Lin, Ying Sheng, Zhanghao Wu, Hao Zhang, Lianmin Zheng, Siyuan Zhuang, Yonghao Zhuang, Joseph E. Gonzalez, Ion Stoica, and Eric P. Xing. 2023. Vicuna: An Open-Source Chatbot Impressing GPT-4 with 90%* ChatGPT Quality. https://lmsys.org/blog/2023-03-30-vicuna/

[8] Jaeyoung Choi, Howard Lei, Venkatesan Ekambaram, Pascal Kelm, Luke Gottlieb, Thomas Sikora, Kannan Ramchandran, and Gerald Friedland. 2013. Human vs machine: establishing a human baseline for multimodal location estimation. In *Proceedings of the ACM International Conference on Multimedia*. ACM, New York, NY, USA, 867–876. doi:10.1145/2502081.2502123

[9] Jinhan Choi, Changhoon Oh, Yea-Seul Kim, and Nam Wook Kim. 2023. VisLab: Enabling Visualization Designers to Gather Empirically Informed Design Feedback. In *Proceedings of the ACM Conference on Human Factors in Computing Systems*. ACM, New York, NY, USA, 813:1–813:18. doi:10.1145/3544548.3581132

[10] Victor Dibia. 2023. LIDA: A Tool for Automatic Generation of Grammar-Agnostic Visualizations and Infographics using Large Language Models. In *Proceedings of the Annual Meeting of the Association for Computational Linguistics*. Association for Computational Linguistics, Toronto, Canada, 113–126. doi:10.18653/v1/2023.acl-demo.11

[11] Deep Ganguli and et al. 2022. Predictability and surprise in large generative models. In *Proceedings of the ACM Conference on Fairness, Accountability, and Transparency*. ACM, New York, NY, USA, 1747–1764. doi:10.1145/3531146.3533229

[12] Timnit Gebru, Jamie Morgenstern, Briana Vecchione, Jennifer Wortman Vaughan, Hanna Wallach, Hal Daumé III, and Kate Crawford. 2021. Datasheets for datasets. *Commun. ACM* 64, 12 (2021), 86–92. doi:10.1145/3458723

[13] Sarah Goodwin, Jason Dykes, Sara Jones, Iain Dillingham, Graham Dove, Alison Duffy, Alexander Kachkaev, Aidan Slingsby, and Jo Wood. 2013. Creative User-Centered Visualization Design for Energy Analysts and Modelers. *IEEE Transactions on Visualization and Computer Graphics* 19, 12 (2013), 2516–2525. doi:10.1109/TVCG.2013.145

[14] Lars Grammel, Melanie Tory, and Margaret-Anne Storey. 2010. How Information Visualization Novices Construct Visualizations. *IEEE Transactions on Visualization and Computer Graphics* 16, 6 (2010), 943–952. doi:10.1109/TVCG.2010.164

[15] Marti A. Hearst, James F. Allen, Eric Horvitz, and Curry I. Guinn. 1999. Trends & Controversies: Mixed-initiative interaction. *IEEE Intell. Syst.* 14, 5 (1999), 14–24. doi:10.1109/5254.796083

[16] Benjamin Heinzerling and Kentaro Inui. 2021. Language Models as Knowledge Bases: On Entity Representations, Storage Capacity, and Paraphrased Queries. In *Proceedings of the Conference of the European Chapter of the Association for Computational Linguistics*. Association for Computational Linguistics, Stroudsburg, PA, USA, 1772–1791. doi:10.18653/v1/2021.eacl-main.153

[17] Aspen K. Hopkins, Michael Correll, and Arvind Satyanarayan. 2020. VisuaLint: Sketchy In Situ Annotations of Chart Construction Errors. *Computer Graphics Forum* 39, 3 (2020), 219–228. doi:10.1111/cgf.13975

[18] Eric Horvitz. 1999. Principles of Mixed-Initiative User Interfaces. In *Proceedings of the ACM Conference on Human Factors in Computing Systems*. ACM, New York, NY, USA, 159–166. doi:10.1145/302979.303030

[19] Kevin Hu, Michiel A. Bakker, Stephen Li, Tim Kraska, and César Hidalgo. 2019. VizML: A Machine Learning Approach to Visualization Recommendation. In *Proceedings of the ACM Conference on Human Factors in Computing Systems*. ACM, New York, NY, USA, 128:1–128:12. doi:10.1145/3290605.3300358

[20] Alicia Key, Bill Howe, Daniel Perry, and Cecilia Aragon. 2012. VizDeck: Self-Organizing Dashboards for Visual Analytics. In *Proceedings of the ACM Conference on Management of Data*. ACM, New York, NY, USA, 681–684. doi:10.1145/2213836.2213931

[21] Gordon Kindlmann and Carlos Scheidegger. 2014. An Algebraic Process for Visualization Design. *IEEE Transactions on Visualization and Computer Graphics* 20, 12 (2014), 2181–2190. doi:10.1109/TVCG.2014.2346325

[22] Abe Leite and Saúl A. Blanco. 2020. Effects of Human vs. Automatic Feedback on Students' Understanding of AI Concepts and Programming Style. In *Proceedings of the ACM Technical Symposium on Computer Science Education*. ACM, New York, NY, USA, 44–50. doi:10.1145/3328778.3366921

[23] Haihan Lin, Derya Akbaba, Miriah Meyer, and Alexander Lex. 2023. Data Hunches: Incorporating Personal Knowledge into Visualizations. *IEEE Transactions on Visualization and Computer Graphics* 29, 1 (2023), 504–514. doi:10.1109/TVCG.2022.3209451

[24] Haihan Lin, Maxim Lisnic, Derya Akbaba, Miriah Meyer, and Alexander Lex. 2026. Here's what you need to know about my data: Exploring Expert Knowledge's Role in Data Analysis. *IEEE Transactions on Visualization and Computer Graphics* 32 (2026), 11 pages. to appear.

[25] Yuyu Luo, Xuedi Qin, Nan Tang, and Guoliang Li. 2018. DeepEye: Towards Automatic Data Visualization. In *Proceedings of the IEEE International Conference on Data Engineering*. IEEE, Piscataway, NJ, USA, 101–112. doi:10.1109/ICDE.2018.00019

[26] Jock Mackinlay. 1986. Automating the Design of Graphical Presentations of Relational Information. *ACM Transactions on Graphics* 5, 2 (1986), 110–141. doi:10.1145/22949.22950

[27] Jock Mackinlay, Pat Hanrahan, and Chris Stolte. 2007. Show Me: Automatic Presentation for Visual Analysis. *IEEE Transactions on Visualization and Computer Graphics* 13, 6 (2007), 1137–1144. doi:10.1109/TVCG.2007.70594

[28] Neha Makhija, Mansi Jain, Nikolaos Tziavelis, Laura Di Rocco, Sara Di Bartolomeo, and Cody Dunne. 2020. Loch Prospector: Metadata Visualization for Lakes of Open Data. In *Proceedings of the IEEE Visualization Conference*. IEEE Computer Society, Los Alamitos, CA, USA, 126–130. doi:10.1109/VIS47514.2020.00032

[29] Damien Masson, Sylvain Malacria, Géry Casiez, and Daniel Vogel. 2023. Charagraph: Interactive Generation of Charts for Realtime Annotation of Data-Rich Paragraphs. In *Proceedings of the ACM Conference on Human Factors in Computing Systems*. ACM, New York, NY, USA, 146:1–146:18. doi:10.1145/3544548.3581091

[30] Andrew McNutt, Gordon Kindlmann, and Michael Correll. 2020. Surfacing Visualization Mirages. In *Proceedings of the ACM Conference on Human Factors in Computing Systems*. ACM, New York, NY, USA, 1–16. doi:10.1145/3313831.3376420

[31] Grégoire Mialon, Roberto Dessì, Maria Lomeli, Christoforos Nalmpantis, Ramakanth Pasunuru, Roberta Raileanu, Baptiste Rozière, Timo Schick, Jane Dwivedi-Yu, Asli Celikyilmaz, Edouard Grave, Yann LeCun, and Thomas Scialom. 2023. Augmented Language Models: a Survey. *Transactions on Machine Learning Research* 2023 (2023), 35 pages. https://openreview.net/forum?id=jh7wH2AzKK

[32] Dominik Moritz, Chenglong Wang, Greg L. Nelson, Halden Lin, Adam M. Smith, Bill Howe, and Jeffrey Heer. 2019. Formalizing Visualization Design Knowledge as Constraints: Actionable and Extensible Models in Draco. *IEEE Transactions on Visualization and Computer Graphics* 25, 1 (2019), 438–448. doi:10.1109/TVCG.2018.2865240

[33] Michael J. Muller and Sarah Kuhn. 1993. Participatory design. *Commun. ACM* 36, 6 (June 1993), 24–28. doi:10.1145/153571.255960

[34] Tamara Munzner. 2009. A Nested Model for Visualization Design and Validation. *IEEE Transactions on Visualization and Computer Graphics* 15, 6 (2009), 921–928. doi:10.1109/TVCG.2009.111

[35] OpenAI. 2023. GPT-4 Technical Report. *CoRR* abs/2303.08774 (2023), 100 pages. arXiv:2303.08774 doi:10.48550/arXiv.2303.08774

[36] OpenAI. 2025. ChatGPT API Documentation. https://platform.openai.com/docs/api-reference/introduction. Accessed: 2025-03-15.

[37] Joon Sung Park, Joseph O'Brien, Carrie Jun Cai, Meredith Ringel Morris, Percy Liang, and Michael S. Bernstein. 2023. Generative Agents: Interactive Simulacra of Human Behavior. In *Proceedings of the Annual ACM Symposium on User Interface Software and Technology* (San Francisco, CA, USA).

Association for Computing Machinery, New York, NY, USA, Article 2, 22 pages. doi:10.1145/3586183.3606763

[38] Joon Sung Park, Lindsay Popowski, Carrie Cai, Meredith Ringel Morris, Percy Liang, and Michael S. Bernstein. 2022. Social Simulacra: Creating Populated Prototypes for Social Computing Systems. In *Proceedings of the ACM Symposium on User Interface Software and Technology*. Association for Computing Machinery, New York, NY, USA, Article 74, 18 pages. doi:10.1145/3526113.3545616

[39] Michael Sedlmair, Miriah Meyer, and Tamara Munzner. 2012. Design Study Methodology: Reflections from the Trenches and the Stacks. *IEEE Transactions on Visualization and Computer Graphics* 18, 12 (2012), 2431–2440. doi:10.1109/TVCG.2012.213

[40] Sungbok Shin, Sunghyo Chung, Sanghyun Hong, and Niklas Elmqvist. 2023. A Scanner Deeply: Predicting Gaze Heatmaps on Visualizations Using Crowdsourced Eye Movement Data. *IEEE Transactions on Visualization and Computer Graphics* 29, 1 (2023), 396–406. doi:10.1109/TVCG.2022.3209472

[41] Sungbok Shin, Sanghyun Hong, and Niklas Elmqvist. 2023. Perceptual Pat: A Virtual Human Visual System for Iterative Visualization Design. In *Proceedings of the ACM Conference on Human Factors in Computing Systems*. ACM, New York, NY, USA, 811:1–811:17. doi:10.1145/3544548.3580974

[42] Arjun Srinivasan, Steven M. Drucker, Alex Endert, and John Stasko. 2019. Augmenting Visualizations with Interactive Data Facts to Facilitate Interpretation and Communication. *IEEE Transactions on Visualization and Computer Graphics* 25, 1 (2019), 672–681. doi:10.1109/TVCG.2018.2865145

[43] Yuan Tian, Weiwei Cui, Dazhen Deng, Xinjing Yi, Yurun Yang, Haidong Zhang, and Yingcai Wu. 2025. ChartGPT: Leveraging LLMs to Generate Charts From Abstract Natural Language. *IEEE Transactions on Visualization and Computer Graphics* 31, 3 (2025), 1731–1745. doi:10.1109/TVCG.2024.3368621

[44] Hugo Touvron, Thibaut Lavril, Gautier Izacard, Xavier Martinet, Marie-Anne Lachaux, Timothée Lacroix, Baptiste Rozière, Naman Goyal, Eric Hambro, Faisal Azhar, Aurélien Rodriguez, Armand Joulin, Edouard Grave, and Guillaume Lample. 2023. LLaMA: Open and Efficient Foundation Language Models. *CoRR* abs/2302.13971 (2023), 27 pages. arXiv:2302.13971 doi:10.48550/arXiv.2302.13971

[45] Jarke J. van Wijk. 2005. The Value of Visualization. In *Proceedings of the IEEE Visualization Conference*. IEEE Computer Society, Los Alamitos, CA, USA, 79–86. doi:10.1109/VISUAL.2005.1532781

[46] Jagoda Walny, Christian Frisson, Mieka West, Doris Kosminsky, Søren Knudsen, Sheelagh Carpendale, and Wesley Willett. 2020. Data Changes Everything: Challenges and Opportunities in Data Visualization Design Handoff. *IEEE Transactions on Visualization and Computer Graphics* 26, 1 (2020), 12–22. doi:10.1109/TVCG.2019.2934538

[47] Chenglong Wang, John Thompson, and Bongshin Lee. 2024. Data Formulator: AI-Powered Concept-Driven Visualization Authoring. *IEEE Transactions on Visualization and Computer Graphics* 30, 1 (2024), 1128–1138. doi:10.1109/TVCG.2023.3326585

[48] Jason Wei, Yi Tay, Rishi Bommasani, Colin Raffel, Barret Zoph, Sebastian Borgeaud, Dani Yogatama, Maarten Bosma, Denny Zhou, Donald Metzler, Ed H. Chi, Tatsunori Hashimoto, Oriol Vinyals, Percy Liang, Jeff Dean, and William Fedus. 2022. Emergent Abilities of Large Language Models. *CoRR* abs/2206.07682 (2022), 30 pages. arXiv:2206.07682 doi:10.48550/arXiv.2206.07682

[49] Yuet Ling Wong, Krishna Madhavan, and Niklas Elmqvist. 2018. Towards Characterizing Domain Experts as a User Group. In *Proceedings of the IEEE Evaluation and Beyond - Methodological Approaches for Visualization Workshop*. IEEE Computer Society, Los Alamitos, CA, USA, 1–10. doi:10.1109/BELIV.2018.8634026

[50] Kanit Wongsuphasawat, Dominik Moritz, Anushka Anand, Jock Mackinlay, Bill Howe, and Jeffrey Heer. 2016. Voyager: Exploratory Analysis via Faceted Browsing of Visualization Recommendations. *IEEE Transactions on Visualization and Computer Graphics* 22, 1 (2016), 649–658. doi:10.1109/TVCG.2015.2467191

[51] Kanit Wongsuphasawat, Zening Qu, Dominik Moritz, Riley Chang, Felix Ouk, Anushka Anand, Jock D. Mackinlay, Bill Howe, and Jeffrey Heer. 2017. Voyager 2: Augmenting Visual Analysis with Partial View Specifications. In *Proceedings of the ACM Conference on Human Factors in Computing Systems*. ACM, New York, NY, USA, 2648–2659. doi:10.1145/3025453.3025768

[52] Shishi Xiao, Suizi Huang, Yue Lin, Yilin Ye, and Wei Zeng. 2024. Let the Chart Spark: Embedding Semantic Context into Chart with Text-to-Image Generative Model. *IEEE Transactions on Visualization and Computer Graphics* 30, 1 (2024), 284–294. doi:10.1109/TVCG.2023.3326913

[53] Yucong Ye, Franz Sauer, Kwan-Liu Ma, Konduri Aditya, and Jacqueline Chen. 2020. A User-Centered Design Study in Scientific Visualization Targeting Domain Experts. *IEEE Transactions on Visualization and Computer Graphics* 26, 6 (2020), 2192–2203. doi:10.1109/TVCG.2020.2970525

[54] Rachael Zehrung, Astha Singhal, Michael Correll, and Leilani Battle. 2021. Vis Ex Machina: An Analysis of Trust in Human versus Algorithmically Generated Visualization Recommendations. In *Proceedings of the ACM Conference on Human Factors in Computing Systems*. ACM, New York, NY, USA, Article 602, 12 pages. doi:10.1145/3411764.3445195

[55] Mengyu Zhou, Qingtao Li, Xinyi He, Yuejiang Li, Yibo Liu, Wei Ji, Shi Han, Yining Chen, Daxin Jiang, and Dongmei Zhang. 2021. Table2Charts: Recommending Charts by Learning Shared Table Representations. In *Proceedings of the ACM Conference on Knowledge Discovery & Data Mining*. ACM, New York, NY, USA, 2389–2399. doi:10.1145/3447548.3467279